\documentclass[aps,preprint,showpacs,showkeys,superscriptaddress]{revtex4-1}
\usepackage{amsfonts}
\usepackage{amsmath}
\usepackage{amssymb}
\usepackage{graphicx}

\RequirePackage[colorlinks,citecolor=blue,urlcolor=blue,linkcolor=blue]{hyperref}
\RequirePackage{color}

\begin{document}

\title{Comment on \textquotedblleft Effective of the q-deformed pseudoscalar
magnetic field on the charge carriers in graphene\textquotedblright }
\author{Angel E. Obispo}
\email[E-mail me at: ]{aeovasquez@gmail.com}
\affiliation{Departamento de F\'{\i}sica - CCET, Universidade Federal do
Maranh\~{a}o (UFMA), Campus Universit\'{a}rio do Bacanga, 65080-805, S\~{a}o
Lu\'{\i}s, MA, Brazil}
\author{Gisele B. Freitas}
\email[E-mail me at: ]{giselebosso@uemasul.edu.br}
\affiliation{Centro de Ci\^{e}ncias Exatas, Naturais e Tecnol\'{o}gicas -
CCENT, Universidade Estadual da Regi\~{a}o Tocantina do Maranh\~{a}o
(UEMASUL), R. Godofredo Viana 1300, 65901- 480, Imperatriz - MA, Brazil}
\author{Luis B. Castro}
\email[E-mail me at: ]{luis.castro@pq.cnpq.br, lrb.castro@ufma.br}
\affiliation{Departamento de F\'{\i}sica - CCET, Universidade Federal do
Maranh\~{a}o (UFMA), Campus Universit\'{a}rio do Bacanga, 65080-805, S\~{a}o
Lu\'{\i}s, MA, Brazil}

\begin{abstract}
We point out a misleading treatment in a recent paper published in this
Journal [J. Math. Phys. (2016) 57, 082105] concerning solutions for the
two-dimensional Dirac-Weyl equation with a q-deformed pseudoscalar magnetic
barrier. The authors misunderstood the full meaning of the potential and
made erroneous calculations, this fact jeopardizes the main results in this
system.
\end{abstract}

\maketitle


In \cite{Iranianos}, Eshgi and Mehraban studied the dynamics of the charge
carriers in graphene in presence of a $q$-deformed pseudoscalar magnetic
barrier. Such barrier is represented by a inhomogeneous background magnetic
field which is associated to a vector potential with a hyperbolic profile as
follows
\begin{equation}
\overrightarrow{A}=2B_{0}d\tanh _{q}(x/2d)\text{\^{e}}_{y},  \label{A}
\end{equation}
\noindent where $B_{0}~$and $d~$are constant and \thinspace $q$ is a
deformation parameter. Note that the expression (\ref{A}) is being
characterized by a $q$-deformed hyperbolic functions, which are based on a $
q $-deformation of the usual hyperbolic functions \cite{Arai}, and are
denoted by (we assume $0<q\leqslant 1$)
\begin{equation}
\sinh _{q}a\equiv \frac{\mathrm{e}^{a}-q\mathrm{e}^{-a}}{2}~~~,~\ \ \cosh
_{q}a\equiv \frac{\mathrm{e}^{a}+q\mathrm{e}^{-a}}{2}~~,  \label{1}
\end{equation}
\noindent which are related as $\cosh _{q}^{2}a-\sinh _{q}^{2}a=q$. Thereby,
the $q$-tangent hyperbolic function is defined via a simple analogy to the
usual hyperbolic functions:%
\begin{equation}
\tanh _{q}a\equiv \frac{\sinh _{q}a}{\cosh _{q}a}~\equiv \frac{1}{\coth _{q}a
}~,  \label{2}
\end{equation}
\noindent which can be re-expressed using (\ref{1}) as\
\begin{equation}
\tanh _{q}(a)=\tanh \left( a-\frac{1}{2}\log q\right) ,  \label{2.1}
\end{equation}
\noindent and~whose derivative
\begin{equation}
\frac{d}{da}\tanh _{q}a=\frac{q}{\cosh _{q}^{2}a}=1-\tanh _{q}^{2}a.
\label{3}
\end{equation}
\noindent Such $q$-deformed hyperbolic functions\ were used in \cite
{Iranianos} to obtain what the authors believed\ to be general exact
expressions for the energies and their associated wavefunctions for the
proposed system. With these results, they also address the scattering regime
to calculate the reflection and transmission coefficients by using the
Riemann's equation. Unfortunately, due to a incorrect manipulation of the
expressions (\ref{2}) and\ (\ref{3}) into the Dirac equation, the results
found in \cite{Iranianos} would not be correct.

It is the aim of this Comment, to point out and correct these mistakes. With
that purpose in mind, we will adopt the notation used in \cite{Iranianos}
and we begin with the correct expression for the background magnetic field $
\overrightarrow{B}$\thinspace $=\overrightarrow{\nabla }\times
\overrightarrow{A}=B(x)$\^{e}$_{z}$,$~$given by
\begin{equation}
\overrightarrow{B}\,=B_{0}\frac{q}{\cosh _{q}^{2}(x/2d)}\text{\^{e}}_{z}
\text{.}  \label{4}
\end{equation}

\begin{figure}[ht]
\begin{center}
\includegraphics[width=11cm, angle=0]{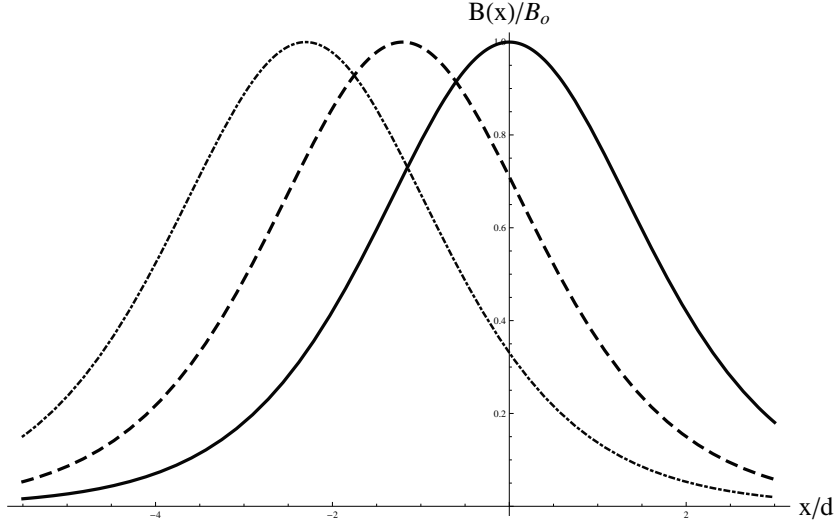}
\end{center}
\par
\vspace*{-0.1cm}
\caption{Plots of the magnetic field $B(x)/B_{0}$ versus $x/2d$, for $q=0.1$
(dot-dashed line), $q=0.3$ (dashed line) and $q=1.0$ (solid line).}
\label{figura1}
\end{figure}

The figure (\ref{figura1}) shows the real behavior of $B(x)/B_{o}~$versus $
x/d~$for different values of $q$. Note that the $q$-parameter does not
deform the magnetic field profile, but only displaces it horizontally. That
behavior is in clear contradiction with the figure (1) in \cite{Iranianos},
where is stated that \textit{the amplitude of magnetic field decrease with
increasing value of }$q$. We disagree with that last statement, due that is
based on incorrect expression for the magnetic field.

Now, to study the dynamics of the carriers charge in graphene in presence of
a background magnetic field, the authors in \cite{Iranianos} used the so-called two-dimensional Dirac-Weyl equation%
\begin{equation}
\overrightarrow{\sigma }.\left[ v_{f}\left( -i\hslash \overrightarrow{\nabla
}+e\overrightarrow{A}\right) \right] \psi (x,y)=E\psi (x,y)  \label{5}
\end{equation}%
\noindent for a given valley degree of freedom. Here, $v_{f}$ $\approx
c/300~ $is the Fermi velocity, $\vec{\sigma}=\left( \sigma _{x},\sigma
_{y}\right) $ are the Pauli matrices and $\psi (x,y)~$is a two-component
spinor, whose transpose is $\psi ^{T}(x,y)=$\textrm{e}$^{ik_{y}y}\left( \psi
_{A}~i\psi _{B}\right) $. The superscripts $A$ and $B$ in the spinor
components designate the triangular sublattices where the electrons are
supported on. The eq. (\ref{5}) represent the version correct of the
Dirac-Weyl equation showed in \cite{Iranianos}, which presents dimensional
inconsistencies that are maintained throughout the paper (see eqs. (1)-(4)
in \cite{Iranianos}).

Substituting $\psi (x,y)$ into equation (\ref{5}), the Weyl-Dirac equation
give rise to two coupled first-order equations for the upper, $\psi _{A}$
and the lower, $\psi _{B}~$components of the spinor%
\begin{eqnarray}
+\left[ \frac{d}{dx}+\left( k_{y}+\frac{e}{\hslash }A_{y}\right) \right]
\psi _{B}(x) &=&\frac{E}{\hslash v_{f}}\psi _{A}(x)\,,  \label{6.1} \\
-\left[ \frac{d}{dx}-\left( k_{y}+\frac{e}{\hslash }A_{y}\right) \right]
\psi _{A}(x) &=&\frac{E}{\hslash v_{f}}\psi _{B}(x)\,.  \label{6.2}
\end{eqnarray}%
\noindent The coupling between the upper and the lower components can be
formally eliminated for $E\neq 0$. Using the expression for
$\psi_{A}$ obtained from (\ref{6.1}) and inserting it in (\ref{6.2}) one
obtains a second-order differential equation for $\psi_{B}$. In a similar
way, using the expression for $\psi_{B}$ obtained from (\ref{6.2}) and
inserting it in (\ref{6.1}) one obtains a second-order differential equation
for $\psi_{A}$. Both results can be written in a compact form:
\begin{equation}
\frac{d^{2}}{dx^{2}}\psi _{\tau }+\left[ \Xi ^{2}-W_{\tau }\right] \psi
_{\tau }=0\,,  \label{7}
\end{equation}%
\noindent where $\tau =\pm 1~$(the upper values$~$correspond to $\psi _{A}~$
and the lower values correspond to $\psi _{B}$),
\begin{equation}
W_{\tau }=\frac{e^{2}}{\hslash ^{2}}A_{y}^{2}+\frac{2e}{\hslash }
k_{y}A_{y}+\tau \frac{e}{\hslash }\frac{dA_{y}}{dx}+k_{y}^{2},  \label{8}
\end{equation}
\noindent and
\begin{equation}
\Xi ^{2}=\frac{E^{2}}{\hslash ^{2}v_{f}^{2}}\,.  \label{9}
\end{equation}%
\noindent These last results tell us that the solutions for this kind of
problem can be formulated as a Sturm--Liouville problem for the component $
\psi _{+}$ and $\psi _{-}$. Nevertheless, the solutions for $E=0$,~excluded
from the Sturm-Liouville problem,$~$was not taken into account in \cite
{Iranianos}. Such solutions (so-called isolated solutions or isolated zero
modes) can be obtained directly form the first--order equations (\ref{6.1})
and (\ref{6.2})
\begin{eqnarray}
\left[ \frac{d}{dx}+\left( k_{y}+\frac{e}{\hslash }A_{y}(x)\right) \right]
\psi _{-} &=&0\,,  \label{10.1} \\
\left[ \frac{d}{dx}-\left( k_{y}+\frac{e}{\hslash }A_{y}(x)\right) \right]
\psi _{+} &=&0\,.  \label{10.2}
\end{eqnarray}
\noindent One can observe that the isolated zero mode for the upper and
lower components are given by
\begin{eqnarray}
\psi _{+}(x) &=&N_{+}\mathrm{e}^{+\mathcal{F}},  \label{11.1} \\
\psi _{-}(x) &=&N_{-}\mathrm{e}^{-\mathcal{F}},  \label{11.2}
\end{eqnarray}
\noindent where $N_{+}$ and $N_{-}$ are normalization constants and
\begin{equation}
\mathcal{F}=\int^{x}d\xi \left( k_{y}+\frac{e}{\hslash }A_{y}(\xi )\right)
\,.  \label{13}
\end{equation}
\noindent In order to guarantee the normalization condition for the zero
mode solutions, the integral must be convergent, i.e.,
\begin{equation}
\int_{-\infty }^{+\infty }dx\left( |N_{+}|^{2}\mathrm{e}^{2\mathcal{F}
}+|N_{-}|^{2}\mathrm{e}^{-2\mathcal{F}}\right) <\infty \,.  \label{14}
\end{equation}
\noindent This result clearly shows that the normalization of the zero mode
is decided by the asymptotic behavior of $\mathcal{F}$. One can check that
it is impossible to have both components different from zero simultaneously
as physically acceptable solutions. So, with the vector potential proposed
in (\ref{A}), the zero mode solutions adopt the explicit form%
\begin{eqnarray}
\psi _{-}^{0} &=&N_{-}\mathrm{e}^{-k_{y}x}\left( \cosh _{q}\frac{x}{2d}
\right) ^{-4d^{2}/l_{B}^{2}}\,,  \label{15.1} \\
\psi _{+}^{0} &=&N_{+}\mathrm{e}^{+k_{y}x}\left( \cosh _{q}\frac{x}{2d}
\right) ^{+4d^{2}/l_{B}^{2}}\,,  \label{15.2}
\end{eqnarray}
\noindent where $l_{B}\equiv \sqrt{\hslash /eB_{0}}~$is the magnetic lenght.
In order to check the normalization condition (\ref{13}), the integral can
be convergent only for $N_{+}=0$ and $2d/l_{B}>k_{y}$. Therefore, the
isolated solution is given by
\begin{equation}
\psi ^{0}(x)=N_{-}\mathrm{e}^{-k_{y}x}\left( \cosh _{q}\frac{x}{2d}\right)
^{-4d^{2}/l_{B}^{2}}\left(
\begin{array}{c}
0 \\
1
\end{array}
\right) \,.  \label{16}
\end{equation}
\noindent With regards to the energy spectrum and the corresponding
eigenstates for $E\neq 0$, the authors in\ \cite{Iranianos} obtained exact
bounded solutions from the second--order differential equations (\ref{7})
with the effective potential $W_{\tau }~$given by eq. (9) in \cite{Iranianos}
. Nevertheless, such potential is dimensionally and structurally wrong, this
due to that the starting point was a incorrect Dirac-Weyl equation (eq. (1)
in \cite{Iranianos}) and also because a careless manipulation of the $q$
-deformed hyperbolic functions. Here we show the correct expression for the
effective potential in the form of a deformed Rosen-Morse potential \cite
{Others,milpas}:
\begin{equation}
W_{\tau }(x)=\frac{1}{l_{B}^{2}}\left( 4\frac{d^{2}}{l_{B}^{2}}-\tau \right)
\tanh _{q}^{2}\left( \frac{x}{2d}\right) +4\frac{k_{y}d}{l_{B}^{2}}\tanh
_{q}\left( \frac{x}{2d}\right) +k_{y}^{2}+\tau \frac{1}{l_{B}^{2}}~.
\label{17}
\end{equation}

\begin{figure}[ht]
\begin{center}
\includegraphics[width=11cm, angle=0]{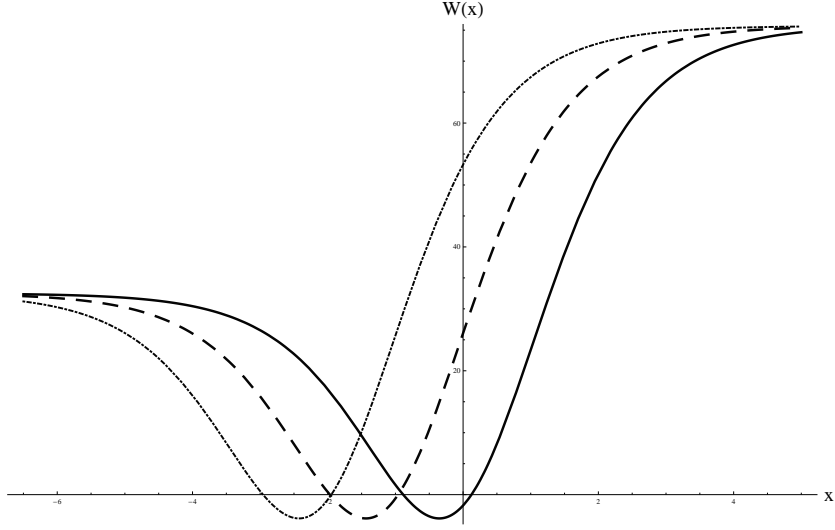}
\end{center}
\par
\vspace*{-0.1cm}
\caption{Plots of the effective potential $W_{-}(x)$ versus $x$, for $q=0.1$
(dot-dashed line), $q=0.3$ (dashed line) and $q=1.0$ (solid line). }
\label{fig2}
\end{figure}

As seen in figure (\ref{fig2}), the potential $W_{\tau }(x)$ for $\tau =-1~$
is characterized for have two maximum values in $W_{-}^{\max }(x\rightarrow
\pm \infty )~=(k_{y}\pm 2d/l_{B}^{2})^{2}~$and one minimum value in $
W_{-}^{\min }=\left( k_{y}^{2}l_{B}^{4}-l_{B}^{2}-4d^{2}\right) /\left(
4d^{2}l_{B}^{2}+l_{B}^{4}\right) $. Note that $W_{-}^{\max }~$and $%
W_{-}^{\min }$ not depend on the deformation parameter $q$, which one,
anecdotally, does not deform the potential and only displaces it
horizontally (the same conclusion was reached for $\tau =+1$). Analitically,
such behavior for $W_{\tau }~$can be proven after replacing (\ref{2.1})\ in (\ref{17}), being now evident that the deformed Rosen-Morse potential depend on the
deformation parameter $q~$only by a translation. In other words, the
parameter $q~$is not necessary to know how many bound-state solutions
exists. This behavior is reflected in the expression for the energy spectrum
\begin{equation}
E_{n}=\pm \frac{\hslash v_{f}}{2d}\sqrt{\left( 8\frac{d^{2}}{l_{B}^{2}}-n\right) n\left[ 1-\left( \frac{2k_{y}d}{n-4d^{2}/l_{B}^{2}}\right) ^{2}\right] },~n=1,2,3...n_{\max }~,  \label{18}
\end{equation}
\noindent which is $q$-independent, as expected. Note that $n~
$and $k_{y}~$are restricted in order to satisfy the square integrability
condition:%
\begin{equation}
n<n_{\max }\left( =4\frac{d^{2}}{l_{B}^{2}}\right) ~~~,~\ \ \left\vert
k_{y}\right\vert \leqslant \frac{1}{2d}\left\vert n-4\frac{d^{2}}{l_{B}^{2}}\right\vert ~.  \label{19}
\end{equation}
\noindent The eigenfunctions associated to (\ref{18}) can be
obtained from the second--order differential equation (\ref{7}) for only one
component of the Dirac spinor, in our case we choose $\psi _{-}~$(=$\psi _{B}$). The
expression for $\psi _{+}~$(=$\psi _{A}$) can be directly built replacing in (\ref{6.1}),
the solution previously obtained for $\psi _{-}~$. In this way, by defining a new
variable $z=\left[ 1+\tanh _{q}\left( x/2d\right) \right] /2$, the general
set complete of solutions can be written as\medskip\
\begin{equation}
\psi ^{E\neq 0}(x,y)=N\mathrm{e}^{-ik_{y}y}(1+z)^{\eta }z^{\mu }\left(
\begin{array}{c}
\frac{i\hslash v_{f}}{Ed}\left[ g(z)F\left( a,b,c;z\right) +\frac{ab}{c}
F\left( a+1,b+1,c+1;z\right) \right] \medskip  \\
F\left( a,b,c;z\right)
\end{array}
\right) ,  \label{20}
\end{equation}
\noindent where $N~$is the normalization constant, $F\left( a,b,c;z\right) ~$%
is the hypergeometric function with
\begin{equation}
a=\mu +\eta -4\frac{d^{2}}{l_{B}^{2}}~~,~~b=\mu +\eta +4\frac{d^{2}}{
l_{B}^{2}}+1~~,\ ~c=2\mu +1,  \label{21}
\end{equation}
\noindent and $g(z)=\left[ \left( \mu -2d^{2}/l_{B}^{2}\right) (1-z)-(\eta
-2d^{2}/l_{B}^{2})z-k_{y}d\right] $ with%
\begin{equation}
\mu =d\sqrt{\left( k_{y}-2\frac{d}{l_{B}^{2}}\right) ^{2}-\Xi ^{2}}~~~,~\ \
\eta =d\sqrt{\left( k_{y}+2\frac{d}{l_{B}^{2}}\right) ^{2}-\Xi ^{2}}.~
\label{22}
\end{equation}
\noindent Normalizable polynomial solutions are obtained by putting $a=-n$,
which allows to rewrite the hypergeometric function $F\left( -n,b,c;z\right)
~$as Jacobi polynomials $P_{n}^{(c-1,-n+b-c)}(z)$. Such mapping is shown in
detail in \cite{milpas}, where the authors also studied the dynamics of the
carriers in graphene subjected to an inhomogeneous magnetic field with a
vector potential $A_{y}=2B_{0}d\tanh (x/2d)$, which is the same from (\ref{A}
) for $q=1$. In such limit, our results are consistent to those found in
\cite{milpas}.

\begin{acknowledgments}
This work was supported in part by means of funds provided by CNPq, Brazil, Grant No. 307932/2017-6 (PQ) and No. 422755/2018-4 (Universal), FAPESP, Brazil, Grant No. 2018/20577-4 and FAPEMA, Brazil, Grant No. UNIVERSAL-01220/18. Angel E. Obispo thanks to CNPq (grant 312838/2016-6) and Secti/FAPEMA (grant FAPEMA DCR-02853/16), for financial support. Gisele B. Freitas also thanks to FAPEMA DCR - 242127/2014.
\end{acknowledgments}

\end{document}